\documentclass[10pt, twocolumn]{IEEEtran}
\usepackage{pifont}
\usepackage{cite}
\usepackage[dvips]{epsfig}
\usepackage{graphicx}
\usepackage{calligra}
\usepackage[T1]{fontenc}
\usepackage{bbold}
\usepackage{mathrsfs}

\usepackage{subfigure}
\usepackage{color}
\usepackage{amsmath}
\usepackage{cases}

\usepackage{amssymb}
\usepackage[justification=centering]{caption}
\usepackage{framed}

\usepackage{subfigure}
\usepackage{color}
\usepackage{amsmath}
\usepackage{bm}
\usepackage{amssymb}
\usepackage{amsbsy}
\usepackage{bbm}
\usepackage{dsfont}
\usepackage[normalem]{ulem}
\usepackage{xcolor,cancel}
\usepackage{enumerate}
\usepackage{paralist}

\usepackage{algorithm, algorithmic}

\usepackage{amsthm}
\usepackage{fancyhdr}

\newtheorem{theorem}{Theorem}

\newtheorem{condition}{Condition}

\newtheorem{corollary}{Corollary}

\newtheorem{definition}{Definition}

\newtheorem{proposition}{Proposition}

\newtheorem{lemma}{Lemma}

\newtheorem{remark}{Remark}

\newtheorem{claim}{Claim}

\newtheorem{assumption}{Assumption}

\newcommand{\ignore}[1]{{}}

\newcounter{parentalgorithm}

\ifCLASSINFOpdf
\else
\fi

\begin{document}
%
\title{Distributed Estimation in Blockchain-aided Internet of Things in the Presence of Attacks}
%
%
%

\author{Hamid Varmazyari, Yiming Jiang, and Jiangfan Zhang, ~\IEEEmembership{Member,~IEEE}
\thanks{H. Varmazyari, Y. Jiang, and J. Zhang are with the Department of Electrical and Computer Engineering, Missouri University of Science and Technology, Rolla MO 65409 USA (e-mail: hvyhw@mst.edu, yjk7z@mst.edu, jiangfanzhang@mst.edu)} 
}



\maketitle
%
\begin{abstract}
Distributed estimation in a blockchain-aided Internet of Things (BIoT) is considered, where the integrated blockchain secures data exchanges across the BIoT and the storage of data at BIoT agents. This paper focuses on developing a performance guarantee for the distributed estimation in a BIoT in the presence of malicious attacks which jointly exploits vulnerabilities present in both IoT devices and the employed blockchain within the BIoT. To achieve this, we adopt the Cramer–Rao Bound (CRB) as the performance metric, and maximize the CRB for estimating the parameter of interest over the attack domain. However, the maximization problem is inherently non-convex, making it infeasible to obtain the globally optimal solution in general. To address this issue, we develop a relaxation method capable of transforming the original non-convex optimization problem into a convex optimization problem. Moreover, we derive the analytical expression for the optimal solution to the relaxed optimization problem. The optimal value of the relaxed optimization problem can be used to provide a valid estimation performance guarantee for the BIoT in the presence of attacks.
\end{abstract}

\begin{IEEEkeywords}
Distributed estimation, Blockchain, double-spending attack, Internet of Things, Cramer-Rao Bound, Water-filling.
\end{IEEEkeywords}

\IEEEpeerreviewmaketitle

\section{Introduction}

The Internet of Things (IoT) encompasses a network of electronic devices, equipped with sensors and software, enabling communication and data exchange through the Internet. As technology advances, adversaries are continuously developing sophisticated techniques to compromise the integrity of IoT networks. Their goals include accessing sensitive information, causing disruption, or even gaining unauthorized control over IoT devices. These threats significantly jeopardize the integrity and security of the data transmitted through IoT devices. Considering the increasing influence of IoT on our daily lives, it is important to address the growing concerns regarding the security issues of IoT\cite{1}.

In a conventional IoT (CIoT) network, IoT devices generate exclusive data pertaining to specific physical phenomena, which are transmitted to a fusion center within the CIoT network for further processing. The CIoT network is vulnerable to various cyber threats. For example, adversaries may intercept data over transmissions or target the fusion center directly, compromising the integrity and confidentiality of the information. Additionally, the reliance on a central authority raises concerns about data privacy and security, since it is susceptible to a single point of failure. These limitations motivate the need for alternative architectures that address the security challenges inherent in CIoT networks. Recently, many research works have delved into the investigation of CIoT vulnerabilities across various applications \cite{2, 3, 4, 5, 6, 7}, ultimately contributing to the enhancement of its security against malicious attacks.

A new technology, blockchain, recently has been integrated into IoT networks. This emergent IoT paradigm is referred to as blockchain-aided IoT (BIoT), which provides a secure way to manage IoT device data in a distributed manner \cite{8, 9}. The BIoT does not rely on an authority to store and process data in a centralized way. Furthermore, the cryptographic techniques of blockchain ensures the security and integrity of the data stored in it. To be specific, each block in a blockchain is linked to the preceding one through a cryptographic hash, establishing a tamper-resistant chain. Additionally, the consensus protocol used in blockchain ensures that every node in the network is able to maintain and synchronize a local copy of the blockchain. Therefore, if the data stored in a node’s local copy of the blockchain is corrupted, it can be easily detected and corrected in the network. These blockchain mechanisms significantly mitigate the vulnerabilities of the IoT network, enhancing its overall security\cite{10, 11, 12, 13, 14, 15}.

Motivated by the benefits of integrating blockchain into IoT, there has been a growing interest in BIoT applications. Some previous literature has studied intrusion detection and anomaly detection in a BIoT network\cite{6, 12, 16, 17}. More recently, numerous studies have investigated estimation problems in BIoT applications, spanning a wide range of domains such as power systems \cite{18, 19}, traffic congestion issues \cite{13}, electric vehicles \cite{20}, unmanned aerial vehicles \cite{21}, and agricultural products \cite{22}. These studies highlight the versatility and applicability of the blockchain-aided IoT in addressing estimation challenges across various sectors. For instance, in power systems, the blockchain-aided IoT can secure the estimation of energy consumption patterns in the presence of attacks, leading to improved resource allocation and grid management. Similarly, in traffic management, estimation algorithms based on blockchain technology can facilitate real-time traffic flow predictions and ensure privacy preservation of users, aiding in congestion mitigation strategies. Furthermore, in domains like electric vehicles and unmanned aerial vehicles, the blockchain-aided IoT can secure data collection and facilitate user authentication, and therefore, enable accurate estimation of battery life or flight paths, optimizing operational efficiency and safety. Additionally, in agriculture, the integration of estimation algorithms with blockchain technology improves crop yield predictions and supply chain management, promoting sustainability and transparency.

Although the BIoT boasts significant security enhancements over the CIoT, it still remains vulnerable to certain attacks targeting the IoT devices and the blockchain within it. Specifically, one of the most devastating threats to blockchain integrity is the double-spending attack (DSA), where adversaries manipulate data stored in existing blocks  of the blockchain employed by a BIoT\cite{23, 24}. Furthermore, if adversaries compromise IoT devices within a BIoT network, they can coerce these devices into generating deceptive and falsified data, and hence may seriously undermine the BIoT\cite{23, 25, 26}.

\subsection{Summary of Results and Main Contributions}
In this paper, we focus on parameter estimation over a BIoT network in the presence of attacks that jointly exploit the vulnerabilities of both IoT devices and the blockchain employed in the BIoT. To be specific, we consider a generic BIoT model stemmed from previous studies on various BIoT applications\cite{13, 15, 16, 17, 18, 19, 20, 21, 22, 30, 31, 32}. In this BIoT model, IoT devices generate data whose distribution is parameterized by a parameter of interest, and the IoT device data are recorded in a blockchain. The estimators in the BIoT utilize the data stored in the blockchain to estimate the parameter of interest. Additionally, we consider an adversary model against the BIoT where an adversary not only attempts to hijack IoT devices but also conducts a DSA on the blockchain within the BIoT. 
In the presence of these attacks, we are interested in the estimation performance guarantee for the BIoT. We adopt the Cramer-Rao Bound (CRB) as the performance metric, and we develop a performance guarantee by maximizing the CRB for estimating the parameter of interest over the attack domain. However, this maximization problem turns out to be non-convex, and hence it is infeasible to obtain its globally optimal solution in general. In order to address this challenge, we develop a relaxation method to convert the non-convex optimization problem into a convex optimization problem. Furthermore, we derive an analytic expression for the optimal solution to the relaxed convex optimization problem, and develop a variant of water-filling procedure to calculate the optimal solution. Our numerical results show that the optimal value of the relaxed optimization problem can be used to provide a valid estimation performance guarantee for the BIoT in the presence of attacks.

\subsection{Related Work}
Distributed estimation in a CIoT in the presence of various types of attacks has been extensively studied in recent literatures see \cite{3, 4, 5, 7, 33, 34, 35} for instance. For example, in \cite{34}, the authors utilize the CRB as the performance metric and propose maximising the CRB for distributed estimation over a CIoT network. However, it is essential to note that the system and adversary models discussed in these papers differ significantly from those considered in this paper. This paper specifically focuses on BIoT rather than CIoT, with a particular emphasis on the vulnerability of the blockchain employed by the BIoT.

Recently, there has been a considerable interest in integrating blockchain technology into various IoT applications. This integration aims to enhance data security of the BIoT applications in the presence of attacks. This trend is evidenced by a plethora of works, including \cite{13, 18, 19, 20, 21, 30, 31, 32, 36, 37, 38, 39, 40}. For example, in \cite{39}, a blockchain-based dynamic topology awareness approach tailored for power distribution networks is developed. This approach is based on a framework of distributed state estimation which can discern topology changes from cyber anomalies in a secure and privacy-protecting manner. In \cite{13}, a peer-to-peer vehicle network is considered, where traffic information is stored in a blockchain to secure information sharing. In \cite{40}, the paper delves into critical challenges associated with state estimation in decentralized power systems. Primarily, it places emphasis on addressing cyber-security concerns within the realm of distributed state estimation. The proposed solution centers around implementing a blockchain framework to bolster the security of the system's database and communication channels. In \cite{18}, the author proposes the integration of distributed state estimation with a blockchain-aided communication platform to secure data transmission and increase the reliability of a power system. In addition, \cite{19} integrates a blockchain into smart grids to secure distributed dynamic state estimation for wide-area smart grids using Kalman filter techniques. However, the proposed method lacks a performance guarantee for the distributed estimation and does not consider attacks on the integrated blockchain. It is worth pointing out that blockchain is assumed perfectly secure in these papers. However, the stored data in the blockchain may not be completely immune to attacks. In contrast, we take the vulnerability of blockchain into account, and analyze the estimation performance of BIoT in the presence of attack that not only target IoT devices but also the blockchain employed by the BIoT.

The paper is organized as follows. In Section \ref{Section_systemmodel}, we describe the BIoT estimation network and the adversary model. Section \ref{Section_Probability_DSA} investigates the estimation performance guarantee for the BIoT over the attack domain. Section \ref{Section_simulation} presents numerical simulations, and our conclusions are provided in Section \ref{Section_conclusion}.
\section{BLOCKCHAIN-AIDED IoT ESTIMATION NETWORK AND ADVERSARY MODELS}\label{Section_systemmodel}
In this section, we introduce a generic BIoT estimation network model that is stemmed from the models employed in the previous studies on BIoT applications \cite{10, 13, 15, 26, 30, 31, 32, 36, 37}. Additionally, we introduce the adopted adversary model which attacks both the IoT devices and the blockchain within a BIoT. The adopted BIoT and adversary models also have been considered in the previous works such as\cite{26}.

\subsection{Blockchain-Aided IoT Estimation Network Model} 
\label{Section_FLBM}
There are mainly three types of agents in a BIoT estimation network: IoT devices (or "things"), miners, and estimators. The first type of agent, IoT devices, embedded with sensors, communication modules, and software, actively produces and exchanges data across a BIoT. The second type of agent, miners, wields substantial memory and computational power, responsible for creating blocks for a blockchain that stores the data generated by IoT devices. The third type of agent, estimators, aims to make accurate and consistent estimates of a parameter of interest based on the data stored within their local blockchain copies. It is worth mentioning that within a BIoT, any agent maintaining a local blockchain copy can act as an estimator, and the consistency of their estimates can be guaranteed by the blockchain consensus protocol. This allows a single agent to simultaneously fulfill different roles. For instance, a miner can also function as an estimator. Considering the widespread adoption of the Proof-of-Work (PoW) consensus protocol \cite{26, 29, 41}, we focus on the BIoT which employs the PoW consensus protocol.

\subsubsection{Data Model} Let $S_{N}\triangleq\{1, 2, ..., N \}$ and $N$ denote the set and the number of IoT devices within a BIoT which sequentially make measurements of a physical phenomenon. Let $\boldsymbol{x}_{j,l}$ denote the measurement made at the $j$-th IoT device in the $l$-th measurement sampling interval. The $j$-th IoT device processes the raw measurement $\boldsymbol{x}_{j,l}$ to produce a summary ${u}_{j,l}$ by employing a function $\mathcal{Q}_j(.)$, that is, ${u}_{j,l}$ = $\mathcal{Q}_j(\boldsymbol{x}_{j,l})$ $\in\mathcal{O} \triangleq \{0, ..., |\mathcal{O}|-1\}$ , where $\mathcal{O}$ is the alphabet set of $u_{j,l}$ with cardinality equal to $|\mathcal{O}|$ for each $j$ and $l$. We assume that the data $\{u_{j,l}\}_{j,l}$ are independent and identically distributed (i.i.d.) and follow the probability mass function (pmf) $\boldsymbol{p}_{\theta}=[p_{\theta,0}, ... , p_{\theta,\mid \mathcal{O}\mid-1}]^T$ which is parameterized by an unknown  parameter $\theta$ of interest .

\subsubsection{Data Exchanges}\label{Section_Block_Mining} The $j$-th IoT device transmits its data $u_{j,l}$, along with its index $l$, to the miners within the BIoT during the $l$-th measurement sampling interval. In every data transformation, a secure hash algorithm (SHA) is employed to encrypt the transmitted data, similar to PoW blockchains \cite{29, 41}. To be specific, before transmitting the data $u_{j,l}$ from the $j$-th IoT device to a miner during the $l$-th measurement sampling interval, the $j$-th IoT device processes its message containing the data $u_{j,l}$ and the data index $l$ by using an SHA, obtaining a message digest. Subsequently, the message digest is encrypted by using a digital signature algorithm with the $j$-th IoT device's private key\cite{26}.

Upon receiving the data, the miner recovers the received message digest by decrypting the digital signature using the public key of the $j$-th IoT device. Simultaneously, the miner generates another message digest by processing the received message with the same SHA. If these two message digests perfectly match, the received message is verified and utilized in subsequent processes. Otherwise, the received message is discarded and a re-transmission occurs\cite{9, 26}.

\subsubsection{Block Mining}
The miners in a BIoT generate blocks to form a chronologically linked blockchain. This process is called mining. Upon receiving and validating messages \(\{u_{j,l}\}\), $\forall j\in S_{N}$ from all IoT devices in the $l$-th measurement sampling interval, each miner constructs a candidate block. This involves aggregating \(\{u_{j,l}\}\), $\forall j\in S_{N}$, the data index \(l\), the digital signatures of IoT device messages, and a header within a block. The block header includes a timestamp, a difficulty value, the hash of the last block (parent block) in the longest branch of the miner’s local copy of the blockchain, a Merkle Root (resulting from recursively hashing data pairs), and a nonce. The hash value of the parent block establishes a cryptographic link between the new block and its predecessor \cite{42}.

Every miner attempts to find its nonce for its candidate block, aiming to achieve a hash value below a predefined difficulty threshold, known as the PoW puzzle. Upon successfully solving the PoW puzzle, the miner broadcasts its candidate block to the other miners and the estimators in the BIoT.
Then the other miners and the estimators carry out a verification and validation process for this candidate block. Specifically, the other miners and the estimators 
first check whether the hash value of the received candidate block falls below the difficulty threshold, confirming that the sender has indeed solved the PoW puzzle. Then, the other miners and estimators scrutinize whether the data index of the received candidate block is only one greater than that of its parent block, establishing the chronological order and integrity of the blockchain.

Once the verification and validation is complete, the other miners and the estimators append the received candidate block after its parent in their local copies of the blockchain. They then shift their focus to mining the subsequent block. This process ensures the security and the integrity of the blockchain by enforcing consensus among network participants before incorporating new blocks into the blockchain.

We assume that each estimator only makes its estimation until the number of blocks in the longest branch of its local blockchain copy reaches a designated threshold \(L\) so that the estimator has enough data to guarantee its estimation performance to meet a prescribed requirement.
\subsection{Adversary Model}
\label{Section_Adversarymodel}

With the intention of misleading the estimators into making erroneous estimations, an adversary attacks both IoT devices and the blockchain in the BIoT to falsify the data used by the estimators.
\subsubsection{Attacks on IoT Devices}
An adversary can attempt to hijack IoT devices and force them to transmit falsified data to miners. Once an adversary controls an IoT device, the adversary can use the IoT device's private key to generate and transmit valid falsified data packages to miners. These falsified data packages can pass the blockchain's authenticity verification process and hence can deceive agents in the BIoT into accepting them as valid. If  the falsified data packages are included in the blocks of the longest branch of the blockchain, they become part of the main chain of the blockchain, and therefore will be used by the estimators to make their estimations.

\subsubsection{Attacks on Blockchain}
An adversary can manipulate the data which have already been stored in a blockchain by creating valid counterfeit blocks. These counterfeit blocks are used to construct a counterfeit branch within the blockchain, with the goal of surpassing the length of the authentic branch. This kind of attack is referred to as DSA. In the context of the BIoT estimation network considered in this paper, a DSA is deemed successful if the counterfeit branch reaches a predefined length $L$ before the authentic branch. Once this occurs, the counterfeit branch becomes the main chain of the blockchain when estimators make their estimations. Therefore, the estimates produced by the estimators rely on the falsified data stored in the counterfeit branch, which result in a significant risk of inaccuracies and errors in the estimations, potentially undermining the integrity and reliability of the BIoT estimation network.

In this paper, we consider an adversary model that jointly attacks both IoT devices and the blockchain within a BIoT. We assume that within the BIoT, an adversary controls certain miners, referred to as malicious miners, and commands them to infiltrate IoT devices via the Internet. Meanwhile, honest miners, the miners not under the adversary's control, actively participate in the mining process, generating blocks to securely store data from IoT devices and form an authentic branch within the blockchain.
At a time instant ${t}_0$, we assume that a subset $\mathcal C_{a}$ $\subset$\ $S_{N}$ of IoT devices are hijacked by the adversary, referred to as malicious IoT devices. Let $\mathcal C_{0} \triangleq\ $$S_{N}$ $\setminus\ \mathcal C_{a} $ represent the set of honest IoT devices. We also assume that at ${t}_0$, the honest miners have generated an authentic branch with ${L}_0\ ({L}_0 \geq 0)$ blocks. Starting from ${t}_0$, the adversary commands the malicious IoT devices to send valid falsified data to miners. Consequently, any data from the malicious IoT devices stored in the blocks of the authentic branch with indices greater than ${L}_0$ are falsified. In order to falsify the data from the malicious IoT devices that have already been stored in the blocks of the authentic branch with indices smaller than or equal to ${L}_0$, the adversary shifts its focus to launching a double-spending attack on the blockchain. Specifically, the adversary allocates all its resources to generate a counterfeit branch within the blockchain, where the data from the malicious IoT devices are falsified. We assume that the counterfeit branch built by the adversary starts from the ${L}_a$-th (${L}_a$ $\leq$\ ${L}_0$)  block of the authentic branch. 


We assume that if the $j$-th IoT device is malicious and $u_{j,l}$ is falsified to $\widetilde u_{j,l}$ by the adversary, then $\widetilde u_{j,l}$ is independent and statistically distributed and follows a malicious-data pmf $\boldsymbol{\tilde{p}}_{\boldsymbol\eta}=[\tilde{p}_{\boldsymbol\eta,0}, \tilde{p}_{\boldsymbol\eta,1}, ... , \tilde{p}_{\boldsymbol\eta,\mid \mathcal{O}\mid-1}]^T$ where $\boldsymbol{{\eta}}\triangleq [{\theta},\boldsymbol{\xi}^T]^T$ and $\boldsymbol{\xi}$ is an unknown attack parameter vector employed by the adversary. The notation $\boldsymbol{\tilde{p}}_{\boldsymbol\eta}$ implies that $\boldsymbol{\tilde{p}}_{\boldsymbol\eta}$ is not only parameterized by the parameter $\theta$ of interest, but also by the attack parameter vector ${\boldsymbol\xi}$. Thus we can express $\boldsymbol{\tilde{p}}_{\boldsymbol\eta}$ as ${\boldsymbol{\tilde{p}}}_{\eta}$ = ${\boldsymbol g}$$({{\theta}}$, ${\boldsymbol{\xi}})$ where ${\boldsymbol {g(.)}}$ represents the functional form of  ${\boldsymbol{\tilde{p}}}_{{\boldsymbol\eta}}$ that depends on how the adversary manipulates the data of the malicious IoT devices. As such, the adversary can influence the estimates produced by the estimators in mainly two ways. Firstly, the adversary can deliberately choose the functional  form ${\boldsymbol {g(.)}}$ of 
 ${\boldsymbol{\tilde{p}}}_{\boldsymbol\eta}$ and the attack parameter vector ${\boldsymbol{\xi}}$ to mislead their estimates. Secondly, the adversary can strategically select the value of $L_a$, which determines the starting block of the double-spending attack, to influence the statistical model of the data stored in the blockchain and consequently impact the estimation performance of the BIoT. 
 \section{Estimation Performance Guarantee for
the BIoT under malicious Attacks}\label{Section_Probability_DSA}
In this section, our focus is on the parameter estimation over a BIoT. We adopt the CRB as the performance metric, which provides a lower but asymptotically achievable bound on the mean squared error of any unbiased estimator \cite{34,43}. We aim to develop an estimation performance guarantee for the BIoT under attacks that jointly exploit the vulnerabilities of the BIoT's blockchain and IoT devices.

It can be shown that the probability of an adversary successfully executing a DSA is a function of $L_0$, $L_a$, and $L$ \cite{25, 26}. We use $P{(L_a)}$ to denote this probability. Let $\hat{u}_{j,l}$ represent the data from the $j$-th IoT device stored in the $l$-th block of the main chain of the blockchain when the main chain reaches a length  of $L$ blocks. If $\hat{u}_{j,l}$ is not falsified by the adversary, then $\hat{u}_{j,l}$ = $u_{j,l}$ which follows the pmf $\boldsymbol{p}_{\theta}=[p_{\theta,0}, ... , p_{\theta,\mid \mathcal{O}\mid-1}]^T$. On the other hand, if $\hat{u}_{j,l}$ is maliciously falsified by the adversary, then $\hat{u}_{j,l}$ = $\tilde{u}_{j,l}$ follows the malicious-data pmf $\boldsymbol{\tilde{p}}_{\eta}=[\tilde{p}_{\eta,0}, ... , \tilde{p}_{\eta,\mid \mathcal{O}\mid-1}]^T$.

When the main chain reaches a length of ${L}$ blocks, let $\hat{\boldsymbol u}\triangleq [ \hat{u}_{1,1}, \hat{u}_{1,2}, \ldots, \hat{u}_{1,L}$
$, \hat{u}_{2,1}, \ldots, \hat{u}_{N,L}]^T$ denote all the IoT device data stored in the main chain. We define $\varphi (\boldsymbol{r}, \boldsymbol{\eta}) \triangleq \Pr {\left\{\hat{\boldsymbol{u}}\ = \boldsymbol{r} \,|\,\boldsymbol{\eta}\right\}}$ 
where $\boldsymbol{r}\triangleq[ {r}_{1,1}, {r}_{1,2} ,\ldots, {r}_{1,L}, {r}_{2,1}, \ldots, {r}_{N,L}]^T$ and ${r}_{j,l} \in \mathcal{O}$ for all $j$ and $l$. Also $\mathcal{R}$ $\triangleq$ $\{\boldsymbol{r}^{1}, \boldsymbol{r}^{2}, \ldots, \boldsymbol{r}^{|\mathcal{R}|} \}$ denotes the set of all possible $\boldsymbol{r}$. Let $\omega$ stand for the event that the DSA is
successful, and $\omega^C$ denotes the event that the DSA is not successful. Then we have
\begin{align}\notag 
&\Pr {\left\{\hat{\boldsymbol{u}}\ = \boldsymbol{r} \,|\,\boldsymbol{\eta}\right\}} \\\notag
&= \Pr {\left\{\hat{\boldsymbol{u}}\ = \boldsymbol{r} \,|\,{\boldsymbol{\eta}, {\omega}}\right\}} \times \Pr {\left\{\omega|{\boldsymbol{\eta} }\right\}}\\\notag
&\quad+ \Pr {\left\{\hat{\boldsymbol{u}}\ = \boldsymbol{r} \,|\,{\boldsymbol{\eta}, \omega^C}\right\}} \times \Pr {\left\{{\omega^C}|{\boldsymbol{\eta} }\right\}}\\\notag
&= {P{(L_a)}} \biggl(\prod_{j\in \mathcal C_{0}} \prod_{l=1}^{L} \Pr {\left\{\hat{u}_{j,l}\ = {r}_{j,l} \,|\,{\boldsymbol{\eta}, {\omega}}\right\}}\\\notag
&\quad\times \prod_{j\in \mathcal C_{a}} \prod_{l=1}^{L} \Pr {\left\{\hat{u}_{j,l}\ = {r}_{j,l} \,|\,{\boldsymbol{\eta}, {\omega}}\right\}}\biggr)\\\notag
&\quad+\Biggl(1-P{(L_a)}\Biggr)\biggl(\prod_{j\in \mathcal C_{0}} \prod_{l=1}^{L} \Pr {\left\{\hat{u}_{j,l}\ = {r}_{j,l} \,|\,{\boldsymbol{\eta}, {\omega}^C}\right\}}\\ 
&\quad\times \prod_{j\in \mathcal C_{a}} \prod_{l=1}^{L} \Pr {\left\{\hat{u}_{j,l}\ = {r}_{j,l} \,|\,{\boldsymbol{\eta}, {\omega}^C}\right\}}\biggr),
\end{align} 
due to the fact that $\{\hat{u}_{j,l}\}$ are statically independent.

If the $j$-th IoT device is honest, then $\hat{u}_{j,l}$ = ${u}_{j,l}$, $\forall l\in \{1,2,\ldots, L\}$. On the other hand, if the $j$-th IoT device is malicious, then $\hat{u}_{j,l}$ = ${u}_{j,l}$, $\forall l\in \{1,2,\ldots, L_{a}-1\}$ and $\hat{u}_{j,l}$ = ${\tilde u}_{j,l}$, $\forall l\in \{L_{0}+1,L_{0}+2,\ldots, L\}$. Moreover, if the DSA launched by the adversary is successful, i.e. $\omega$ happens, then $\hat{u}_{j,l}$ = ${\tilde u}_{j,l}$, $\forall l\in \{L_{a},L_{a}+1,\ldots, L_{0}\}$. Otherwise, $\hat{u}_{j,l}$ = ${u}_{j,l}$, $\forall l\in \{L_{a},L_{a}+1,\ldots, L_{0}\}$. As such, we can obtain
\begin{align}\notag
    &\prod_{j\in \mathcal C_{0}} \prod_{l=1}^{L} \Pr {\left\{\hat{u}_{j,l}\ = {r}_{j,l} \,|\,{\boldsymbol{\eta}, {\omega}}\right\}}\\\notag
    &\times \prod_{j\in \mathcal C_{a}} \prod_{l=1}^{L} \Pr {\left\{\hat{u}_{j,l}\ = {r}_{j,l} \,|\,{\boldsymbol{\eta}, {\omega}}\right\}} \\\notag
\end{align}
\begin{align}
    &= \prod_{j\in \mathcal C_{0}} \prod_{l=1}^{L} p_{\theta,r_{j,l}} \prod_{j\in \mathcal C_{a}} \prod_{l=1}^{L_a-1} p_{\theta,r_{j,l}} \prod_{l=L_a}^{L_0} \tilde{p}_{\boldsymbol\eta,r_{j,l}} \prod_{l=L_0+1}^{L} \tilde{p}_{\boldsymbol\eta,r_{j,l}}.
\end{align}
Similar to (2), we can obtain
\begin{align}\notag
    &\prod_{j\in \mathcal C_{0}} \prod_{l=1}^{L} \Pr {\left\{\hat{u}_{j,l}\ = {r}_{j,l} \,|\,{\boldsymbol{\eta}, {\omega}^C}\right\}}\\\notag
    &\times \prod_{j\in \mathcal C_{a}} \prod_{l=1}^{L} \Pr {\left\{\hat{u}_{j,l}\ = {r}_{j,l} \,|\,{\boldsymbol{\eta}, {\omega}^C}\right\}}\\
    &= \prod_{j\in \mathcal C_{0}} \prod_{l=1}^{L} p_{\theta,r_{j,l}}\prod_{j\in \mathcal C_{a}} \prod_{l=1}^{L_a-1} p_{\theta,r_{j,l}} \prod_{l=L_a}^{L_0} p_{\theta,r_{j,l}} \prod_{l=L_0+1}^{L} \tilde{p}_{\boldsymbol\eta,r_{j,l}}.
\end{align}
From (1), (2), and (3), we can show that the distribution $\varphi (\boldsymbol{r}, \boldsymbol{\eta})$ of $\hat{\boldsymbol{u}}$ can be expressed as  
\begin{align}\notag
&\varphi (\boldsymbol{r},\boldsymbol{\eta}) \\\notag
&=\prod_{j\in \mathcal C_{0}} \prod_{l=1}^{L} p_{\theta,r_{j,l}}\\\notag
&\quad\times\Biggl\{ {P{(L_a)}}\Biggl(\prod_{j\in \mathcal C_{a}} \prod_{l=1}^{L_a-1} p_{\theta,r_{j,l}} \prod_{l=L_a}^{L_0} \tilde{p}_{\boldsymbol\eta,r_{j,l}} \prod_{l=L_0+1}^{L} \tilde{p}_{\boldsymbol\eta,r_{j,l}}\Biggr)\\\notag
&\quad + \Biggl(1-P{(L_a)}\Biggr)\\
&\quad\times\Biggl(\prod_{j\in \mathcal C_{a}} \prod_{l=1}^{L_a-1} p_{\theta,r_{j,l}} \prod_{l=L_a}^{L_0} p_{\theta,r_{j,l}} \prod_{l=L_0+1}^{L} \tilde{p}_{\boldsymbol\eta,r_{j,l}}\Biggr)\Biggr\}.
\end{align}
For the sake of notational simplicity, we define
\begin{align}   
\varphi_{0} (\boldsymbol{r}) \triangleq \prod_{j\in \mathcal C_{0}} \prod_{l=1}^{L} p_{\theta,r_{j,l}},
\end{align}
and
\begin{align}\notag
 &\varphi_{a} (\boldsymbol{r},\boldsymbol{\eta}) \\\notag
&\triangleq {P{(L_a)}}\times\Biggl(\prod_{j\in \mathcal C_{a}} \prod_{l=1}^{L_a-1} p_{\theta,r_{j,l}} \prod_{l=L_a}^{L_0} \tilde{p}_{\boldsymbol\eta,r_{j,l}} \prod_{l=L_0+1}^{L} \tilde{p}_{\boldsymbol\eta,r_{j,l}}\Biggr)\\\notag
&\quad + \Biggl(1-P{(L_a)}\Biggr)\\
&\quad\times\Biggl(\prod_{j\in \mathcal C_{a}} \prod_{l=1}^{L_a-1} p_{\theta,r_{j,l}} \prod_{l=L_a}^{L_0} p_{\theta,r_{j,l}} \prod_{l=L_0+1}^{L} \tilde{p}_{\boldsymbol\eta,r_{j,l}}\Biggr),
\end{align}
for any $\boldsymbol{r} \in \mathcal{R}$.
Hence, $\varphi (\boldsymbol{r},\boldsymbol{\eta})$ can be rewritten as
\begin{align}
    \varphi (\boldsymbol{r},\boldsymbol{\eta}) = \varphi_{0} (\boldsymbol{r})\varphi_{a} (\boldsymbol{r},\boldsymbol{\eta}).
\end{align}
It is worth mentioning that the parameter $\theta$ of interest and the attack parameter vector $\boldsymbol{\xi}$ are both unknown to the estimators. The estimators should jointly estimate both of them, that is, $\boldsymbol{\eta}$. The Fisher Information Matrix (FIM) $\boldsymbol{J}_{\boldsymbol{\eta}}$ for estimating $\boldsymbol{\eta}$ is defined as \cite{43}
\begin{align}
\left[\boldsymbol{J}_{\boldsymbol{\eta}}\right]_{l,m} &\triangleq -E\left[\frac{\partial^2{L}(\boldsymbol{r},\boldsymbol{\eta})}{\partial \eta_l \partial \eta_m}\right],
\end{align}
where $\left[\boldsymbol{J}_{\boldsymbol{\eta}}\right]_{l,m}$ denotes the element in the $l$-th row and the $m$-th column of the matrix $\boldsymbol{J}_{\boldsymbol{\eta}}$. The term ${L}(\boldsymbol{r},\boldsymbol{\eta})$ in (8) is the log-likelihood function, which can be expressed as
\begin{align}\notag
{L}(\boldsymbol{r},\boldsymbol{\eta}) &= \ln {\Pr\left\{\hat{\boldsymbol{u}} = \boldsymbol{r} \,|\, \boldsymbol{\eta}\right\}}\\\notag
&= \sum_{\boldsymbol{r} \in {\mathcal{R}}} \mathbbm{1}{\left\{ \hat{\boldsymbol{u}} = \boldsymbol{r} \right\}} \ln {\varphi(\boldsymbol{r},\boldsymbol{\eta})}\\
&= \sum_{\boldsymbol{r} \in {\mathcal{R}}} \mathbbm{1}{\left\{ \hat{\boldsymbol{u}} = \boldsymbol{r} \right\}}\biggl[\ln{\varphi_{0} (\boldsymbol{r}) + \ln{\varphi_{a} (\boldsymbol{r},\boldsymbol{\eta})}\biggr]},
\end{align}
where $\mathbbm{1}{\left\{ \hat{\boldsymbol{u}} = \boldsymbol{r} \right\}}$ stands for the indicator function that $\mathbbm{1}{\left\{ \hat{\boldsymbol{u}} = \boldsymbol{r} \right\}}$ = 1 if $\hat{\boldsymbol{u}} = \boldsymbol{r}$ and 0 otherwise\cite{4}.
The first order derivative of ${L}(\boldsymbol{r},\boldsymbol{\eta})$ can be expressed as
\begin{align}\notag
\frac{\partial{L}(\boldsymbol{r},\boldsymbol{\eta})}{\partial \boldsymbol{\eta}} &= \begin{bmatrix}
\frac{\partial {L}(\boldsymbol{r},\boldsymbol{\eta})}{\partial{\theta}} \\
\frac{\partial {L}(\boldsymbol{r},\boldsymbol{\eta})}{\partial \boldsymbol{\xi}}
\end{bmatrix}
\\
&= \begin{bmatrix}
&\sum_{\boldsymbol{r}\in \mathcal{R}} \frac{1}{\varphi_{{0}} (\boldsymbol{r})} \frac{\partial \varphi_{0} (\boldsymbol{r})}{\partial {\theta}} \mathbbm{1} {\left\{ \hat{\boldsymbol{u}} = \boldsymbol{r} \right\}}\\
&+\sum_{\boldsymbol{r}\in \mathcal{R}} \frac{1}{{\varphi_a(\boldsymbol{r},\boldsymbol{\eta})}} \frac{\partial {\varphi_a(\boldsymbol{r},\boldsymbol{\eta})}}{\partial {\theta}} \mathbbm{1}{\left\{ \hat{\boldsymbol{u}} = \boldsymbol{r} \right\}} \\
&\sum_{\boldsymbol{r}\in \mathcal{R}} \frac{1}{\varphi_{{0}} (\boldsymbol{r})} \frac{\partial \varphi_{0} (\boldsymbol{r})}{\partial \boldsymbol{\xi}} \mathbbm{1} {\left\{ \hat{\boldsymbol{u}} = \boldsymbol{r} \right\}} \\
&+ \sum_{\boldsymbol{r}\in \mathcal{R}} \frac{1}{{\varphi_a(\boldsymbol{r},\boldsymbol{\eta})}} \frac{\partial {\varphi_a(\boldsymbol{r},\boldsymbol{\eta})}}{\partial \boldsymbol{\xi}} \mathbbm{1}{\left\{ \hat{\boldsymbol{u}} = \boldsymbol{r} \right\}} 
\end{bmatrix},
\end{align}
where $\frac{\partial{L}(\boldsymbol{r},\boldsymbol{\eta})}{\partial \boldsymbol{\xi}}$ is a vector. We know from (5) that $\frac{\partial \varphi_{0} (\boldsymbol{r})}{\partial \boldsymbol{\xi}}$ = \textbf{0} since $\varphi_{0} (\boldsymbol{r})$ does not depend on $\boldsymbol{\xi}$.  As a result, we have $\sum_{\boldsymbol{r}\in \mathcal{R}} \frac{1}{\varphi_{0} (\boldsymbol{r})} \frac{\partial \varphi_{0} (\boldsymbol{r})}{\partial \boldsymbol{\xi}} \mathbbm{1} {\left\{\hat{\boldsymbol{u}} = \boldsymbol{r} \right\}} = \textbf{0}$, and therefore, (10) can be simplified to
\begin{align}
\frac{\partial{L}(\boldsymbol{r},\boldsymbol{\eta})}{\partial \boldsymbol{\eta}}
&= \begin{bmatrix}
&\sum_{\boldsymbol{r}\in \mathcal{R}} \frac{1}{\varphi_{{0}} (\boldsymbol{r})} \frac{\partial \varphi_{0} (\boldsymbol{r})}{\partial {\theta}} \mathbbm{1} {\left\{ \hat{\boldsymbol{u}} = \boldsymbol{r} \right\}}\\
&+\sum_{\boldsymbol{r}\in \mathcal{R}} \frac{1}{{\varphi_a(\boldsymbol{r},\boldsymbol{\eta})}} \frac{\partial {\varphi_a(\boldsymbol{r},\boldsymbol{\eta})}}{\partial {\theta}} \mathbbm{1}{\left\{ \hat{\boldsymbol{u}} = \boldsymbol{r} \right\}} \\
&\sum_{\boldsymbol{r}\in \mathcal{R}} \frac{1}{{\varphi_a(\boldsymbol{r},\boldsymbol{\eta})}} \frac{\partial {\varphi_a(\boldsymbol{r},\boldsymbol{\eta})}}{\partial \boldsymbol{\xi}} \mathbbm{1}{\left\{ \hat{\boldsymbol{u}} = \boldsymbol{r} \right\}}
\end{bmatrix}.
\end{align}
Furthermore, the second order derivative of ${L}(\boldsymbol{r},\boldsymbol{\eta})$ can be written as
\begin{align}
&\frac{\partial^2{{L}}(\boldsymbol{\eta})}{\partial \boldsymbol{\eta}^2} =
  \begin{bmatrix}
 \frac{\partial^2 {L}(\boldsymbol{r},\boldsymbol{\eta})}{\partial {\theta}^2} & (\frac{\partial^2 {L}(\boldsymbol{r},\boldsymbol{\eta})}{\partial {\theta}\partial\boldsymbol{\xi}})^T \\
\frac{\partial^2 {L}(\boldsymbol{r},\boldsymbol{\eta})}{\partial\boldsymbol{\xi}\partial{\theta}} & \frac{\partial^2 {L}(\boldsymbol{r},\boldsymbol{\eta})}{\partial \boldsymbol{\xi}^2} 
\end{bmatrix},
\end{align}
where 
\begin{align}\notag
\frac{\partial^2 {L}(\boldsymbol{r},\boldsymbol{\eta})}{\partial{\theta}^2} &= \sum_{\boldsymbol{r}\in \mathcal{R}} \Biggr\{\frac{-1}{\varphi_0^2(\boldsymbol{r})} {\biggr[\frac{\partial \varphi_{0} (\boldsymbol{r})}{\partial {\theta}}\biggl]}^2\\\notag
&\quad+ \frac{1}{\varphi_{{0}} (\boldsymbol{r})} \frac{\partial^2\varphi_{0} (\boldsymbol{r})}{\partial{\theta}^2}\Biggl\} \mathbbm{1} {\left\{ \hat{\boldsymbol{u}} = \boldsymbol{r} \right\}} 
\\\notag
&\quad+ \sum_{\boldsymbol{r}\in \mathcal{R}} \Biggr\{\frac{-1}{\varphi_a^2(\boldsymbol{r},\boldsymbol{\eta})} {\biggr[\frac{\partial {\varphi_a(\boldsymbol{r},\boldsymbol{\eta})}}{\partial {\theta}}\biggl]}^2\\
&\quad+ \frac{1}{\varphi_a(\boldsymbol{r},\boldsymbol{\eta})} \frac{\partial^2 {\varphi_a(\boldsymbol{r},\boldsymbol{\eta})}}{\partial {\theta}^2}\Biggl\} \mathbbm{1}{\left\{ \hat{\boldsymbol{u}} = \boldsymbol{r} \right\}},
\end{align}
\begin{align}\notag
\frac{\partial^2 {L}(\boldsymbol{r},\boldsymbol{\eta})}{\partial {\theta}\partial\boldsymbol{\xi}} &= \frac{\partial^2 {L}(\boldsymbol{r},\boldsymbol{\eta})}{\partial\boldsymbol{\xi}\partial {\theta}} \\\notag
&= \sum_{\boldsymbol{r}\in \mathcal{R}} \Biggr\{\frac{-1}{\varphi_a^2(\boldsymbol{r},\boldsymbol{\eta})} \frac{\partial {\varphi_a(\boldsymbol{r},\boldsymbol{\eta})}}{\partial {\theta}}{\biggr[\frac{\partial {\varphi_a(\boldsymbol{r},\boldsymbol{\eta})}}{\partial \boldsymbol{\xi}}\biggl]} \\
&\quad+ \frac{1}{\varphi_a(\boldsymbol{r},\boldsymbol{\eta})} \frac{\partial^2{\varphi_a(\boldsymbol{r},\boldsymbol{\eta})}}{\partial {\theta}\partial\boldsymbol{\xi}}\Biggl\} \mathbbm{1} {\left\{ \hat{\boldsymbol{u}} = \boldsymbol{r} \right\}}, 
\end{align}
and
\begin{align}\notag
\frac{\partial^2 {L}(\boldsymbol{r},\boldsymbol{\eta})}{\partial \boldsymbol{\xi}^2} &= \sum_{\boldsymbol{r}\in \mathcal{R}} \Biggr\{\frac{-1}{\varphi_a^2(\boldsymbol{r},\boldsymbol{\eta})} {\biggr[\frac{\partial {\varphi_a(\boldsymbol{r},\boldsymbol{\eta})}}{\partial \boldsymbol{\xi}}\biggl]\biggr[\frac{\partial {\varphi_a(\boldsymbol{r},\boldsymbol{\eta})}}{\partial \boldsymbol{\xi}}\biggl]^T} \\ &\quad+ \frac{1}{\varphi_a(\boldsymbol{r},\boldsymbol{\eta})} \frac{\partial^2{\varphi_a(\boldsymbol{r},\boldsymbol{\eta})}}{\partial \boldsymbol{\xi}^2}\Biggl\} \mathbbm{1}{\left\{ \hat{\boldsymbol{u}} = \boldsymbol{r} \right\}}.
\end{align}
Note that by interchanging the derivative and the summation, we can obtain
\begin{align}
    &\sum_{\boldsymbol{r}\in \mathcal{R}}\frac{1}{\varphi_{{0}} (\boldsymbol{r})} \frac{\partial^2\varphi_{0} (\boldsymbol{r})}{\partial{\theta}^2}=\frac{1}{\varphi_{{0}} (\boldsymbol{r})} \frac{\partial^2\sum_{\boldsymbol{r}\in \mathcal{R}}\varphi_{0} (\boldsymbol{r})}{\partial{\theta}^2}= 0.
\end{align}
By employing (16), we can simplify (13), (14), and (15) to
\begin{align}\notag
    \frac{\partial^2 {L}(\boldsymbol{r},\boldsymbol{\eta})}{\partial{\theta}^2} &= \sum_{\boldsymbol{r}\in \mathcal{R}} \Biggr\{\frac{-1}{\varphi_0^2(\boldsymbol{r})} {\biggr[\frac{\partial \varphi_{0} (\boldsymbol{r})}{\partial {\theta}}\biggl]}^2 \Biggl\} \mathbbm{1} {\left\{ \hat{\boldsymbol{u}} = \boldsymbol{r} \right\}} 
\\
&\quad+\sum_{\boldsymbol{r}\in \mathcal{R}} \Biggr\{\frac{-1}{\varphi_a^2(\boldsymbol{r},\boldsymbol{\eta})} {\biggr[\frac{\partial {\varphi_a(\boldsymbol{r},\boldsymbol{\eta})}}{\partial {\theta}}\biggl]}^2 \Biggl\} \mathbbm{1}{\left\{ \hat{\boldsymbol{u}} = \boldsymbol{r} \right\}},
\end{align}
\begin{align}\notag 
    &\frac{\partial^2 {L}(\boldsymbol{r},\boldsymbol{\eta})}{\partial {\theta}\partial\boldsymbol{\xi}} \\
    &= \sum_{\boldsymbol{r}\in \mathcal{R}} \Biggr\{\frac{-1}{\varphi_a^2(\boldsymbol{r},\boldsymbol{\eta})} \frac{\partial {\varphi_a(\boldsymbol{r},\boldsymbol{\eta})}}{\partial {\theta}}{\biggr[\frac{\partial {\varphi_a(\boldsymbol{r},\boldsymbol{\eta})}}{\partial \boldsymbol{\xi}}\biggl]} \Biggl\} \mathbbm{1} {\left\{ \hat{\boldsymbol{u}} = \boldsymbol{r}\right\}},
\end{align}
\begin{align}\notag
    &\frac{\partial^2 {L}(\boldsymbol{r},\boldsymbol{\eta})}{\partial \boldsymbol{\xi}^2} \\
&= \sum_{\boldsymbol{r}\in \mathcal{R}} \Biggr\{\frac{-1}{\varphi_a^2(\boldsymbol{r},\boldsymbol{\eta})} {\biggr[\frac{\partial {\varphi_a(\boldsymbol{r},\boldsymbol{\eta})}}{\partial \boldsymbol{\xi}}\biggl]\biggr[\frac{\partial {\varphi_a(\boldsymbol{r},\boldsymbol{\eta})}}{\partial \boldsymbol{\xi}}\biggl]^T} \Biggl\} \mathbbm{1}{\left\{ \hat{\boldsymbol{u}} = \boldsymbol{r} \right\}}.
\end{align}

By employing (17)--(19) and the fact that $E\{\mathbbm{1}{\left\{ \hat{\boldsymbol{u}} = \boldsymbol{r} \right\}}\} = \varphi (\boldsymbol{r}, \boldsymbol{\eta}) = {\varphi_{0}} (\boldsymbol{r}){\varphi}_{a} (\boldsymbol{r},\boldsymbol{\eta})$, we can obtain
\begin{align}\notag
 {\boldsymbol{J}}_{\boldsymbol{\xi}} &\triangleq -E\left\{ \frac{\partial^2 {L}(\boldsymbol{r},\boldsymbol{\eta})}{\partial {\boldsymbol{\xi}}^2} \right\}
 \\
&=\sum_{\boldsymbol{r}\in \mathcal{R}}\frac{{\varphi}_{0} (\boldsymbol{r})}{\varphi_{a}(\boldsymbol{r}, \boldsymbol\eta)}  \left(\frac{\partial\varphi_{a}(\boldsymbol r, \boldsymbol\eta)}{\partial\boldsymbol{\xi}}\right)\biggr(\frac{\partial {\varphi_a(\boldsymbol{r},\boldsymbol{\eta})}}{\partial \boldsymbol{\xi}}\biggl)^T,
\end{align}
\begin{align}\notag
 \boldsymbol{f}_{a} &\triangleq -E\left\{\frac{\partial^2 {L}(\boldsymbol{r},\boldsymbol{\eta})}{\partial{\theta}{\partial{\boldsymbol{\xi}}}}\right\} \\
 &=  \sum_{\boldsymbol{r}\in \mathcal{R}}\frac{{\varphi}_{0}(\boldsymbol{r})}{\varphi_{a}(\boldsymbol{r}, \boldsymbol\eta)}\frac{\partial\varphi_{a}(\boldsymbol r, \boldsymbol\eta)}{\partial{\theta}} \left(\frac{\partial\varphi_{a}(\boldsymbol r, \boldsymbol\eta)}{\partial{\boldsymbol{\xi}}}\right),
 \end{align}
\begin{align}
&{J}_{\theta} \triangleq -E\left\{\frac{\partial^2 {L}(\boldsymbol{r},\boldsymbol{\eta})}{\partial{\theta}^2} \right\} =
   {J}_{\mathcal C_{0}} + {J}_{\mathcal C_{a}},
\end{align}
where
\begin{align}
 &{{J}}_{\mathcal C_{a}} \triangleq \sum_{\boldsymbol{r}\in \mathcal{R}}\frac{{\varphi}_{0}(\boldsymbol{r})}{\varphi_{a}(\boldsymbol{r}, \boldsymbol\eta)} \left(\frac{\partial\varphi_{a}(\boldsymbol r, \boldsymbol\eta)}{\partial{\theta}}\right)^2,
 \end{align}
 \begin{align}
 &{{J}}_{\mathcal C_{0}} \triangleq \sum_{\boldsymbol{r}\in \mathcal{R}}\frac{{\varphi}_{a}(\boldsymbol{r},\boldsymbol{\eta})}{\varphi_{0}(\boldsymbol{r})}  \left(\frac{\partial\varphi_{0}(\boldsymbol r)}{\partial{\theta}}\right)^2.
\end{align}
Moreover, the first order derivatives of ${\varphi_{0}} (\boldsymbol{r})$ and ${\varphi}_{a} (\boldsymbol{r},\boldsymbol{\eta})\ $in (20)--(24) can be expressed as
\begin{align}
\frac{\partial\varphi_{0}(\boldsymbol r)}{\partial{\theta}} = \sum_{j\in \mathcal C_{0}}\sum_{l=1}^{L} \biggl(\frac{\partial p_{\theta,r_{j,l}} }{\partial{\theta}} \prod_{k \in \{ 1, \ldots, L\} \setminus \{l\}} p_{\theta,r_{j,k}}\biggr),
\end{align}
\begin{align}\notag
& \frac{\partial\varphi_{a}(\boldsymbol r, \boldsymbol\eta)}{\partial\boldsymbol{\xi}} = P{(L_a)}\Biggl(\sum_{j\in \mathcal C_{a}}\Biggl\{\Biggl[\sum_{l=L_{a}}^{L} \biggl(\frac{\partial \tilde{p}_{\boldsymbol\eta,r_{j,l}} }{\partial\boldsymbol{\xi}}\\\notag
&\times\prod_{k \in \{ L_{a}, \ldots, L_0, L_0+1, \ldots, L\} \setminus \{l\}} \tilde{p}_{\boldsymbol\eta,r_{j,k}}\biggr)\Biggr] \prod_{l=1}^{L_{a}-1} p_{\theta,r_{j,l}}\Biggr\}\Biggr)\\\notag
& +\Biggl(1-P{(L_a)}\Biggr) \Biggl(\sum_{j\in \mathcal C_{a}} \Biggl\{\Biggl[\sum_{l=L_0+1}^{L} \biggl(\frac{\partial \tilde{p}_{\boldsymbol\eta,r_{j,l}} }{\partial\boldsymbol{\xi}}\\
&\times\prod_{k \in \{  L_0+1, \ldots, L\} \setminus \{l\}} \tilde{p}_{\boldsymbol\eta,r_{j,k}}\biggr)\Biggr] \prod_{l=1}^{L_a-1} p_{\theta,r_{j,l}}\prod_{l=L_a}^{L_0} p_{\theta,r_{j,l}}\Biggr\}\Biggr),
\end{align}
\begin{align}\notag
&\frac{\partial\varphi_{a}(\boldsymbol r, \boldsymbol\eta)}{\partial{\theta}} = P{(L_a)}\Biggl(\sum_{j\in \mathcal C_{a}}\Biggl\{ \Biggl[\sum_{l=1}^{L_a-1} \biggl(\frac{\partial p_{\theta,r_{j,l}} }{\partial{\theta}}\\\notag
&\times\prod_{k \in \{ 1, \ldots, L_a-1\} \setminus \{l\}} p_{\theta,r_{j,k}}\biggr)\Biggr] \prod_{l=L_a}^{L_0} \tilde{p}_{\boldsymbol\eta,r_{j,l}} \prod_{l=L_0+1}^{L} \tilde{p}_{\boldsymbol\eta,r_{j,l}}\\\notag
& + \Biggl[\sum_{l=L_{a}}^{L} \biggl(\frac{\partial \tilde{p}_{\boldsymbol\eta,r_{j,l}} }{\partial{\theta}}\\\notag
&\times\prod_{k \in \{ L_{a}, \ldots, L_0, L_0+1, \ldots, L\} \setminus \{l\}} \tilde{p}_{\boldsymbol\eta,r_{j,k}}\biggr)\Biggr] \prod_{l=1}^{L_{a}-1} p_{\theta,r_{j,l}}\Biggr\}\Biggr)\\\notag
& +\Biggl(1-P{(L_a)}\Biggr) \Biggl(\sum_{j\in \mathcal C_{a}} \Biggl\{ \Biggl[\sum_{l=1}^{L_0} \biggl(\frac{\partial p_{\theta,r_{j,l}} }{\partial{\theta}}\\\notag
&\times\prod_{k \in \{  1, \ldots, L_a-1, L_a, \ldots, L_0\} \setminus \{l\}} p_{\theta,r^i_{j,k}}\biggr)\Biggr] \prod_{l=L_0+1}^{L} \tilde{p}_{\boldsymbol\eta,r_{j,l}}\\\notag
&+ \Biggl[\sum_{l=L_0+1}^{L} \biggl(\frac{\partial \tilde{p}_{\boldsymbol\eta,r_{j,l}} }{\partial{\theta}}\\
&\times\prod_{k \in \{  L_0+1, \ldots, L\} \setminus \{l\}} \tilde{p}_{\boldsymbol\eta,r_{j,k}}\biggr)\Biggr] \prod_{l=1}^{L_a-1} p_{\theta,r_{j,l}}\prod_{l=L_a}^{L_0} p_{\theta,r_{j,l}}\Biggr\}\Biggr).
\end{align}
According to the definition of the FIM in (8), we can obtain
\begin{align}
\boldsymbol{J}_{\boldsymbol{\eta}} =
\begin{bmatrix}
{J}_{\theta} & {\boldsymbol{f}}^T_{a}  \\
{\boldsymbol{f}}_{a} & {\boldsymbol{J}}_{\boldsymbol{\xi}}
\end{bmatrix},
\end{align}
by employing (20)--(24). Notice that the CRB${_{\theta}}$ for estimating ${\theta}$ is defined as $[\boldsymbol{J}_{\boldsymbol{\eta}}^{-1}]_{1,1}$ which can be written as
\begin{align}\notag
    {\text{CRB}}{_{\theta}}&\triangleq [\boldsymbol{J}_{\boldsymbol{\eta}}^{-1}]_{1,1}\\\notag
    &=\left( {J}_{\theta} - {\boldsymbol{f}}^T_{a}{\boldsymbol{J}}^{-1}_{\boldsymbol{\xi}}{\boldsymbol{f}}_{a} \right)^{-1} \\
    &= \biggl[{J}_{\mathcal C_{0}} +\left({J}_{\mathcal C_{a}}-{\boldsymbol{f}}^T_{a}{\boldsymbol{J}}^{-1}_{\boldsymbol{\xi}}{\boldsymbol{f}}_{a}\right)\biggr]^{-1}.
\end{align}
As shown in (20)--(24) and (29), the adversary can deliberately choose the functional form ${\boldsymbol {g(.)}}$ of ${\boldsymbol{\tilde{p}}}_{{\boldsymbol\eta}}$ and the attack parameter vector ${\boldsymbol{\xi}}$ to affect CRB$_{{\theta}}$. Moreover, the adversary can strategically select the value of ${L_a}$ to degrade CRB$_{{\theta}}$. In light of these observations, an estimation performance guarantee for estimating ${\theta}$ can be developed by maximizing CRB$_{{\theta}}$ over ${\boldsymbol {g}}$, ${\boldsymbol{\xi}}$, and ${L_a}$,
which can be cast as the following optimization problem:

\begin{align}
\tag{30a}
\max_{L_a} \max_{\{\boldsymbol{g},{\boldsymbol{\xi}}\}}\quad &\biggl[{J}_{\mathcal C_{0}} +({J}_{\mathcal C_{a}}-{\boldsymbol{f}}^T_{a}{\boldsymbol{J}}^{-1}_{\boldsymbol{\xi}}{\boldsymbol{f}}_{a})\biggr]^{-1}\\
\tag{30b}
 s.t.\quad \ \ & \sum^{|\mathcal{R}|}_{i=1}[{\boldsymbol g}({{\theta}}, {\boldsymbol{\xi}})]_i = 1,\\
\tag{30c}
&0 \leq[{\boldsymbol g}({{\theta}}, {\boldsymbol{\xi}})]_i\leq 1,  \ \ \  \forall i=1, 2, \ldots, |\mathcal{R}|,
\end{align}
where $[{\boldsymbol g}({{\theta}}, {\boldsymbol{\xi}})]_i$ denotes the $i$-th element of the vector ${\boldsymbol g}({{\theta}}, {\boldsymbol{\xi}})$.
It can be shown that the optimization problem in (30) is non-convex, and hence it is infeasible to obtain its optimal solution in general. To address this issue, we first develop an achievable upper bound on the objective function in (30a). 

We define 
a row vector $\boldsymbol \gamma_{\theta}$ and a matrix $\boldsymbol{\Xi}_{\boldsymbol{\xi}}$ as follows
\begin{align}
\tag{31}
\boldsymbol \gamma_{\theta}\triangleq \biggl[\psi^\theta_{0}, \psi^\theta_{1}, \ldots, \psi^\theta_{|\mathcal{R}|-1}\biggr] \ \text{and}\  \boldsymbol{\Xi}_{\boldsymbol{\xi}}\triangleq \biggl[\boldsymbol\psi^{\boldsymbol{\xi}}_{0}, \boldsymbol\psi^{\boldsymbol{\xi}}_{1}, \ldots, \boldsymbol\psi^{\boldsymbol{\xi}}_{|\mathcal{R}|-1}\biggr],
\end{align}
where the scalar ${\psi^{\theta}_{i}}$ and the vector ${\boldsymbol\psi^{\boldsymbol{\xi}}_{i}}$ in (31) are defined as

\begin{align}\notag
&{\psi^{\theta}_{i}} \triangleq \sqrt\frac{{{\varphi}_{0} (\boldsymbol{r}^i)}}{\varphi_{a}(\boldsymbol r^i, \boldsymbol\eta)}\frac{\partial\varphi_{a}(\boldsymbol r^i, \boldsymbol\eta)}{\partial{\theta}}, \ \ \forall i=1, 2, \ldots, |\mathcal{R}|, \\
\tag{32}
&{\boldsymbol\psi^{\boldsymbol{\xi}}_{i}} \triangleq \sqrt\frac{{{\varphi}_{0} (\boldsymbol{r}^i)}}{\varphi_{a}(\boldsymbol r^i, \boldsymbol\eta)}\frac{\partial\varphi_{a}(\boldsymbol r^i, \boldsymbol\eta)}{\partial{\boldsymbol{\xi}}}, \ \ \forall i=1, 2, \ldots, |\mathcal{R}|,
\end{align}
and $\boldsymbol r^i$ is the $i$-th element of $\mathcal{R}$ = $\{\boldsymbol{r}^{1}, \boldsymbol{r}^{2}, \ldots, \boldsymbol{r}^{|\mathcal{R}|} \}$. By employing (31), ${\boldsymbol{J}}_{\boldsymbol{\xi}}$, ${{\boldsymbol{f}}_{a}}$, and ${J}_{\mathcal C_{a}}$ in (20), (21) and (23) can be rewritten in compact forms as
\begin{align}
\tag{33}
&{\boldsymbol{J}}_{\boldsymbol{\xi}} =  \boldsymbol{\Xi}_{\boldsymbol{\xi}}\boldsymbol{\Xi}_{\boldsymbol{\xi}}^T,\\
\tag{34}
&{{\boldsymbol{f}}_{a}} = \boldsymbol{\Xi}_{\boldsymbol{\xi}}\boldsymbol \gamma_{\theta}^T,\\
\tag{35}
&{J}_{\mathcal C_{a}} = \boldsymbol \gamma_{\theta}\boldsymbol \gamma_{\theta}^T.
\end{align}
We have the following theorem regarding the objective function in (30a).
  
 $Theorem\ 1$: The CRB for estimating $\theta$ is bounded above as per 
 \begin{align}
 \tag{36}
    &{\text{CRB}}{_{\theta}} = \biggl[{J}_{\mathcal C_{0}} +\left({J}_{\mathcal C_{a}}-{\boldsymbol{f}}^T_{a}{\boldsymbol{J}}^{-1}_{\boldsymbol{\xi}}{\boldsymbol{f}}_{a}\right)\biggr]^{-1} \leq {{J}_{\mathcal C_{0}}^{-1}}.
 \end{align}
Equality in (36) holds if and only if 
$\boldsymbol \gamma_{\theta}$ lies in the span of the rows of  $\boldsymbol{\Xi}_{\boldsymbol{\xi}}$.

\begin{IEEEproof}    
By employing (33), (34), and (35), we can obtain
\begin{align}\notag
   &{{J}}_{\mathcal C_{a}}-\boldsymbol{{f}}^T_{a}\boldsymbol{{J}}^{-1}_{\boldsymbol{\xi}}\boldsymbol{{f}}_{a}\\\notag&=\boldsymbol \gamma_{\theta}\boldsymbol \gamma_{\theta}^T - \boldsymbol \gamma_{\theta}\boldsymbol{\Xi}_{\boldsymbol{\xi}}^T \biggr(\boldsymbol{\Xi}_{\boldsymbol{\xi}}\boldsymbol{\Xi}_{\boldsymbol{\xi}}^T\biggl)^{-1} \boldsymbol{\Xi}_{\boldsymbol{\xi}}\boldsymbol \gamma_{\theta}^T
   \\\notag& =\boldsymbol \gamma_{\theta}\boldsymbol \gamma_{\theta}^T - \boldsymbol \gamma_{\theta}\boldsymbol{\Xi}_{\boldsymbol{\xi}}^T\boldsymbol{h} - \boldsymbol{h}^T\boldsymbol{\Xi}_{\boldsymbol{\xi}}\boldsymbol \gamma_{\theta}^T + \boldsymbol{h}^T\boldsymbol{\Xi}_{\boldsymbol{\xi}}\boldsymbol{\Xi}_{\boldsymbol{\xi}}^T\boldsymbol{h} \\
   \tag{37}
   &= \biggr(\boldsymbol \gamma_{\theta}^T - {\boldsymbol{\Xi}_{\boldsymbol{\xi}}^T\boldsymbol{h}\biggl)}^T \biggr(\boldsymbol \gamma_{\theta}^T - {\boldsymbol{\Xi}_{\boldsymbol{\xi}}^T\boldsymbol{h}\biggl)}
   \geq 0,
\end{align}
where $\boldsymbol{h} \triangleq(\boldsymbol{\Xi}_{\boldsymbol{\xi}}\boldsymbol{\Xi}_{\boldsymbol{\xi}}^T)^{-1} \boldsymbol{\Xi}_{\boldsymbol{\xi}}\boldsymbol \gamma_{\theta}^T$. The equality in (37) is attained if and only if 
\begin{align}
\tag{38}
\boldsymbol \gamma_{\theta}=\boldsymbol\kappa^T \boldsymbol{\Xi}_{\boldsymbol{\xi}},
\end{align}
for some vector $\boldsymbol\kappa$, which implies $\boldsymbol \gamma_{\theta}$ lies in the span of the rows of  $\boldsymbol{\Xi}_{\boldsymbol{\xi}}$.

Finally, from (29), (37) and (38), we can conclude that 
\begin{align}\notag
    \text{CRB}{_{\theta}} &= \biggr({J}_{{\theta}}-\boldsymbol{f}^T_{a}\boldsymbol{J}^{-1}_{\boldsymbol{\xi}}\boldsymbol{f}_{a}\biggl)^{-1} \\\notag
    &= \biggr[{J}_{\mathcal C_{0}} + \biggr({J}_{\mathcal C_{a}}-\boldsymbol{f}^    T_{a}\boldsymbol{J}^{-1}_{\boldsymbol{\xi}}\boldsymbol{f}_{a}\biggl)\biggl]^{-1}\\&  \leq {{J}_{\mathcal C_{0}}}^{-1},
    \tag{39}
\end{align}
with equality if and only if $\boldsymbol \gamma_{\theta}$ lies in the span of the rows of  $\boldsymbol{\Xi}_{\boldsymbol{\xi}}$.
\end{IEEEproof}
Theorem 1 demonstrates that in order to degrade the CRB for estimating $\theta$ to the largest extent, the adversary needs to guarantee $\boldsymbol \gamma_{\theta}=\boldsymbol\kappa^T \boldsymbol{\Xi}_{\boldsymbol{\xi}}$ for some vector $\boldsymbol\kappa$.
In light of this, the optimization problem in (30) can be simplified to 
\begin{align}
\tag{40a}
\max_{L_a} \max_{\{\boldsymbol{g},{\boldsymbol{\xi}}\}}\quad &\left({J}_{\mathcal C_{0}}\right)^{-1}\\
\tag{40b}
s.t. \quad \ \ &\boldsymbol \gamma_{\theta}=\boldsymbol\kappa^T \boldsymbol{\Xi}_{\boldsymbol{\xi}}, \\
\tag{40c}
& \sum^{|\mathcal{R}|}_{i=1}[{\boldsymbol g}({{\theta}}, {\boldsymbol{\xi}})]_i = 1,\\
\tag{40d}
&0 \leq[{\boldsymbol g}({{\theta}}, {\boldsymbol{\xi}})]_i\leq 1,  \ \ \  \forall i=1, 2, \ldots, |\mathcal{R}|.
\end{align}
To solve the optimization problem in (40), we first define two terms ${X}_{\boldsymbol{r}}$, and $Y_{\boldsymbol{r}}$ as follows
\begin{align}\notag
{X}_{\boldsymbol{r}} &\triangleq \frac{\Biggl[\sum_{j\in \mathcal C_{0}}\sum_{l=1}^{L} \biggl(\frac{\partial p_{\theta,r_{j,l}} }{\partial{\theta}} \prod_{k \in \{ 1, \ldots, L\} \setminus \{l\}} p_{\theta,r_{j,k}}\biggr)\Biggr]^2}{\varphi_{0} (\boldsymbol{r})} \\
\tag{41}
&= \frac{\Biggl[\sum_{j\in \mathcal C_{0}}\sum_{l=1}^{L} \biggl(\frac{\partial p_{\theta,r_{j,l}} }{\partial{\theta}} \prod_{k \in \{ 1, \ldots, L\} \setminus \{l\}} p_{\theta,r_{j,k}}\biggr)\Biggr]^2}{\prod_{j\in \mathcal C_{0}} \prod_{l=1}^{L} p_{\theta,r_{j,l}}},
\end{align}
\begin{align}\notag
Y_{\boldsymbol{r}} &\triangleq P{(L_a)}\Biggl(\prod_{j\in \mathcal C_{a}}\prod_{l=1}^{L_a-1} p_{\theta,r_{j,l}} \prod_{l=L_a}^{L_0} \tilde{p}_{\boldsymbol\eta,r_{j,l}} \prod_{l=L_0+1}^{L} \tilde{p}_{\boldsymbol\eta,r_{j,l}}\Biggr) \\\notag
&\quad+ \Biggl(1-P{(L_a)}\Biggr)\\& \tag{42}
\quad\times\Biggl(\prod_{j\in \mathcal C_{a}} \prod_{l=1}^{L_a-1} p_{\theta,r_{j,l}} \prod_{l=L_a}^{L_0} p_{\theta,r_{j,l}} \prod_{l=L_0+1}^{L} \tilde{p}_{\boldsymbol\eta,r_{j,l}}\Biggr).
\end{align}
From (6) and (42), we know that $Y_{\boldsymbol{r}} = \varphi_{a}(\boldsymbol r, \boldsymbol\eta)$. By employing (41) and (42), $(J_{\mathcal{C}_{0}})^{-1}$ can be rewritten as follows
\begin{align}\notag
&(J_{\mathcal{C}_{0}})^{-1} = \Biggl[\sum_{\boldsymbol{r}\in \mathcal{R}}\\\notag &\left\{\frac{ {P{(L_a)}}\!\Biggl(\!\prod_{j\in \mathcal C_{a}}\! \prod_{l=1}^{L_a-1} p_{\theta,r_{j,l}}\! \prod_{l=L_a}^{L_0} \tilde{p}_{\boldsymbol\eta,r_{j,l}} \!\prod_{l=L_0+1}^{L} \tilde{p}_{\boldsymbol\eta,r_{j,l}}\!\!\Biggr)}{\prod_{j\in \mathcal C_{0}} \prod_{l=1}^{L} p_{\theta,r_{j,l}}} \right.\\\notag
&+\Biggl(1-P{(L_a)}\Biggr)\\\notag&\left.\times\frac{
\Biggl(\prod_{j\in \mathcal C_{a}} \prod_{l=1}^{L_a-1} p_{\theta,r_{j,l}} \prod_{l=L_a}^{L_0} p_{\theta,r_{j,l}} \notag \prod_{l=L_0+1}^{L} \tilde{p}_{\boldsymbol\eta,r_{j,l}}\Biggr)}{\prod_{j\in \mathcal C_{0}} \prod_{l=1}^{L} p_{\theta,r_{j,l}}}\right\}\\\notag 
  &\times \biggl[\sum_{j\in \mathcal C_{0}}\sum_{l=1}^{L} \biggl(\frac{\partial p_{\theta,r_{j,l}} }{\partial{\theta}} \prod_{k \in \{ 1, \ldots, L\} \setminus \{l\}} p_{\theta,r_{j,k}}\biggr)\biggr]^2\Biggr]^{-1} \\
  \tag{43}
  &= \Biggl[\sum_{\boldsymbol{r}\in \mathcal{R}} Y_{\boldsymbol{r}} X_{\boldsymbol{r}} \Biggr]^{-1},
\end{align}
which implies that in order to solve the optimization problem in (40), we can first solve the following optimization problem
\begin{align}
\tag{44a}
\min_{L_a} \min_{\{\boldsymbol{g},{\boldsymbol{\xi}}\}}\qquad &\sum_{\boldsymbol{r}\in \mathcal{R}} Y_{\boldsymbol{r}} X_{\boldsymbol{r}}\\\notag
\tag{44b}s.t. \qquad \ \ &\text{(40b)--(40d)},
\end{align}
and then take the reciprocal of the minimal objective function in (44a) to obtain the optimal value of (40).

Noting that $X_{\boldsymbol{r}}$ is constant, we have
\begin{align}
\tag{45}
    \min_{L_a} \min_{\{\boldsymbol{g},{\boldsymbol{\xi}}\}}\sum_{\boldsymbol{r}\in \mathcal{R}} Y_{\boldsymbol{r}} X_{\boldsymbol{r}} \geq \min_{\{Y_{\boldsymbol{r}}\}_{\boldsymbol{r}\in \mathcal{R}}}\sum_{\boldsymbol{r}\in \mathcal{R}} Y_{\boldsymbol{r}} X_{\boldsymbol{r}}.
\end{align}
Furthermore, it is seen from (6), (7) and (42) that
\begin{align}
\tag{46}
    \sum^{|\mathcal{R}|}_{i=1}[{\boldsymbol g}({{\theta}}, {\boldsymbol{\xi}})]_i = \sum_{\boldsymbol{r}\in \mathcal{R}} \varphi_{0} (\boldsymbol{r})Y_{\boldsymbol{r}} = 1.
\end{align}
Hence, the constraint in (40c) is equivalent to 
\begin{align}
\tag{47}
    \sum_{\boldsymbol{r}\in \mathcal{R}} \varphi_{0} (\boldsymbol{r})Y_{\boldsymbol{r}} = 1.
\end{align}
By dropping the constraint in (40b), the optimization problem in (44) can be relaxed to
\begin{align}
\tag{48a}
    \min_{\{Y_{\boldsymbol{r}}\}_{\boldsymbol{r}\in \mathcal{R}}}\qquad & \sum_{\boldsymbol{r}\in \mathcal{R}} Y_{\boldsymbol{r}} X_{\boldsymbol{r}}\\
    \tag{48b}
    s.t. \qquad \ \ & \sum_{\boldsymbol{r}\in \mathcal{R}} \varphi_{0} (\boldsymbol{r})Y_{\boldsymbol{r}} = 1,\\
    \tag{48c}
    & Y_{\boldsymbol{r}} \in [0,1],\ \ \ \forall \ {\boldsymbol{r}\in \mathcal{R}}.
\end{align}
The optimal solution to (48) can be obtained by solving the Karush-Kuhn-Tucker (KKT) conditions which can be written as
\begin{align}
\tag{49a}
  & \frac{\partial \mathcal{L}}{\partial Y_{\boldsymbol{r}}} = 0, \ \ \ \forall \ {\boldsymbol{r}\in \mathcal{R}},\\
  \tag{49b}
  &\sum_{\boldsymbol{r}\in \mathcal{R}} \varphi_{0} (\boldsymbol{r}) Y_{\boldsymbol{r}} = 1,\\
  \tag{49c}
&\mu_{\boldsymbol{r}} \geq 0,\ \ \ \forall \ {\boldsymbol{r}\in \mathcal{R}},\\
\tag{49d}
&\nu_{\boldsymbol{r}} \geq 0,\ \ \ \forall \ {\boldsymbol{r}\in \mathcal{R}},\\
\tag{49e}
  &\mu_{\boldsymbol{r}}(Y_{\boldsymbol{r}}) = 0,\ \ \ \forall \ {\boldsymbol{r}\in \mathcal{R}},\\\notag
  \tag{49f}
  &\nu_{\boldsymbol{r}}(1 - Y_{\boldsymbol{r}}) = 0, \ \ \ \forall \ {\boldsymbol{r}\in \mathcal{R}},
\end{align}
where
\begin{align}\notag
    \mathcal{L} &\triangleq \sum_{\boldsymbol{r}\in \mathcal{R}} Y_{\boldsymbol{r}} X_{\boldsymbol{r}} - \lambda\left(\sum_{\boldsymbol{r}\in \mathcal{R}} \varphi_{0} (\boldsymbol{r}) Y_{\boldsymbol{r}} - 1\right)\\
    \tag{50}
    &\quad- \sum_{\boldsymbol{r}\in \mathcal{R}} \mu_{\boldsymbol{r}} Y_{\boldsymbol{r}} - \sum_{\boldsymbol{r}\in \mathcal{R}} \nu_{\boldsymbol{r}}(1 - Y_{\boldsymbol{r}}),
\end{align}
and $\lambda$ is the Lagrange multiplier associated with the equality constraint in (48b). The variables $\mu_{\boldsymbol{r}}$ and $\nu_{\boldsymbol{r}}$ are the Lagrange multipliers associated with the lower and upper bound constraints in (48c), respectively.

By employing (50), the equation (49a) can be simplified to
\begin{align}
\tag{51}
    \frac{\partial \mathcal{L}}{\partial Y_{\boldsymbol{r}}} = X_{\boldsymbol{r}} - \lambda \varphi_{0} (\boldsymbol{r}) - \mu_{\boldsymbol{r}} + \nu_{\boldsymbol{r}} = 0.
\end{align}
Let $\nu^{\ast}_{\boldsymbol{r}}$, and $\mu^{\ast}_{\boldsymbol{r}}$ denote the optimal Lagrange multipliers. From (49c)--(49f), we know that the optimal solution $Y^{\ast}_{\boldsymbol{r}}$ satisfies
\begin{subnumcases}{}
    Y^{\ast}_{\boldsymbol{r}} = 0, & if $\mu^{\ast}_{\boldsymbol{r}} > 0, \nu^{\ast}_{\boldsymbol{r}} = 0$, \tag{52} \\
    Y^{\ast}_{\boldsymbol{r}} = 1, & if $\mu^{\ast}_{\boldsymbol{r}} = 0, \nu^{\ast}_{\boldsymbol{r}} > 0$, \tag{53} \\
    Y^{\ast}_{\boldsymbol{r}} \in (0,1), & if $\mu^{\ast}_{\boldsymbol{r}} = \nu^{\ast}_{\boldsymbol{r}} = 0$. \tag{54}
\end{subnumcases}
By employing (52)--(54), (49b), and (51), the closed form expression for the optimal value $Y^{\ast}_{\boldsymbol{r}}$ can be written as
\begin{equation}
\resizebox{\linewidth}{!}{
$Y^{\ast}_{\boldsymbol{r}} = 
\begin{cases}
0 \ \ \ \ \ \ \ \ \ \ \ \ \ \ \ \ \ \forall \ \boldsymbol{r}\in S_1 \triangleq \{\boldsymbol{r}: \lambda < \frac{X_{\boldsymbol{r}}}{\varphi_{0} (\boldsymbol{r})}\}, \\
1 \ \ \ \ \ \ \ \ \ \ \ \ \ \ \ \ \ \forall \ \boldsymbol{r}\in S_2 \triangleq \{\boldsymbol{r}: \lambda > \frac{X_{\boldsymbol{r}}}{\varphi_{0} (\boldsymbol{r})}\}, \\
\frac{1-\sum_{\boldsymbol{r}\in S_2}{\varphi_{0} (\boldsymbol{r})}}{\sum_{\boldsymbol{r}\in S_3}{\varphi_{0} (\boldsymbol{r})}} \ \ \forall \ \boldsymbol{r}\in S_3 \triangleq \{\boldsymbol{r}: \lambda = \frac{X_{\boldsymbol{r}}}{\varphi_{0} (\boldsymbol{r})}\}.
\end{cases}$
} 
\tag{55}
\end{equation}
\begin{figure}[!h]
  \centering
    \includegraphics[width=9cm]{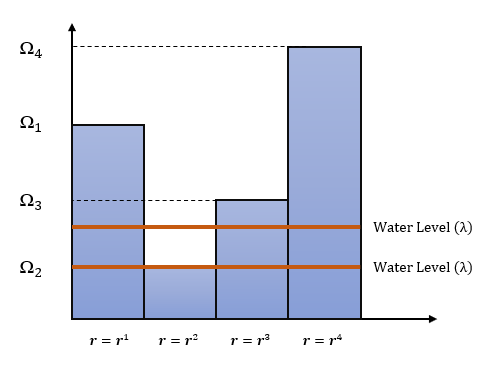}
  \caption{Water-filling method.}
\end{figure}
By evaluating $(\sum_{\boldsymbol{r}\in \mathcal{R}} Y^{\ast}_{\boldsymbol{r}} X_{\boldsymbol{r}})^{-1}$ by using (55), we can obtain an estimation performance guarantee for the BIoT in the presence of attacks.

As demonstrated by (55), deriving the optimal value for $Y^{\ast}_{\boldsymbol{r}}$ necessitates determining the optimal Lagrange multiplier $\lambda^{\ast}$. To tackle this challenge, we propose  a variant of the water-filling procedure. To be specific, we define $\Omega_{\boldsymbol{r}} = \frac{X_{\boldsymbol{r}}}{\varphi_{0} (\boldsymbol{r})}$ for each $\boldsymbol{r}\in \mathcal{R}$, and increase the water-level, i.e., $\lambda$, starting from zero. If $\lambda < \min_{\boldsymbol{r}} (\Omega_{\boldsymbol{r}})$, then $Y^{\ast}_{\boldsymbol{r}} = 0$ for all $\boldsymbol{r}\in \mathcal{R}$. It's evident that substituting $Y^{\ast}_{\boldsymbol{r}} = 0$ for all $\boldsymbol{r}\in \mathcal{R}$ into (49b) renders the constraint in (49b) unsatisfied. On the other hand, if $\lambda > \max_{\boldsymbol{r}} (\Omega_{\boldsymbol{r}})$, then $Y^{\ast}_{\boldsymbol{r}} = 1$ for all $\boldsymbol{r}\in \mathcal{R}$, which also violates in (49b). Thus, the water level $\lambda$ must be in the range $(\min_{\boldsymbol{r}}(\Omega_{\boldsymbol{r}}),\max_{\boldsymbol{r}}(\Omega_{\boldsymbol{r}}))$. In light of this, we only need to increase the water-level $\lambda$ starting from $\min_{\boldsymbol{r}}(\Omega_{\boldsymbol{r}})$ to $\max_{\boldsymbol{r}}(\Omega_{\boldsymbol{r}})$. During the process of increasing the water level $\lambda$, we evaluate $Y^{\ast}_{\boldsymbol{r}}$ for each $\boldsymbol{r}$ by using (55) and verify the constraint in (49b). Specifically, if $\lambda > \Omega_{\boldsymbol{r}}$, then $Y^{\ast}_{\boldsymbol{r}} = 1$. If $\lambda < \Omega_{\boldsymbol{r}}$, then $Y^{\ast}_{\boldsymbol{r}} = 0$, and if $\lambda = \Omega_{\boldsymbol{r}}$, then we need to check if $Y^{\ast}_{\boldsymbol{r}} = \frac{1-\sum_{\boldsymbol{r}\in S_2}{\varphi_{0} (\boldsymbol{r})}}{\sum_{\boldsymbol{r}\in S_3}{\varphi_{0} (\boldsymbol{r})}} \in [0,1]$. If the constraint in (49b) is satisfied by ${\{Y^{\ast}_{\boldsymbol{r}}\}_{\boldsymbol{r}\in \mathcal{R}}}$ and $Y^{\ast}_{\boldsymbol{r}}\in [0,1]$ for all ${\boldsymbol{r}\in \mathcal{R}}$, the water level $\lambda$ stops increasing, and the corresponding ${\{Y^{\ast}_{\boldsymbol{r}}\}_{\boldsymbol{r}\in \mathcal{R}}}$ are the optimal solution to the relaxed optimization problem in (48). Moreover, the optimal objective for the relaxed optimization in (48) can be attained by calculating $\sum_{\boldsymbol{r}\in \mathcal{R}} Y^{\ast}_{\boldsymbol{r}} X_{\boldsymbol{r}}$. If the constraint in (49b) is not satisfied by a water level $\lambda$ or the corresponding $Y^{\ast}_{\boldsymbol{r}}$ is not within the range [0,1] for all ${\boldsymbol{r}\in \mathcal{R}}$, the water level continues to rise until the constraint in (49b) is satisfied and $Y^{\ast}_{\boldsymbol{r}} \in [0,1]$ for all ${\boldsymbol{r}\in \mathcal{R}}$. We elaborate on this water-filling process by using an example where $\boldsymbol{r} \in\mathcal{R} = \{\boldsymbol{r}^1,\boldsymbol{r}^2,\boldsymbol{r}^3,\boldsymbol{r}^4\}$ as illustrated in Fig. 1. The water level $\lambda$ ascends from the minimum value $\Omega_{2}$ to the maximum value $\Omega_{4}$, as the constraint in (49b) cannot be met when $\lambda > \Omega_{4}$ or $\lambda < \Omega_{2}$. Commencing from $\Omega_{2}$, we first determine the value of $Y_{\boldsymbol{r}}$ for each $\boldsymbol{r}$. When $\lambda = \Omega_{2}$, $Y^{\ast}_{1} = Y^{\ast}_{3} = Y^{\ast}_{4} = 0$, and we need to check if $Y^{\ast}_{2} \in [0,1]$. If $Y^{\ast}_{2} \in [0,1]$, and these ${\{Y_{\boldsymbol{r}}\}_{\boldsymbol{r}\in \mathcal{R}}}$ satisfy the constraint in (49b), the optimal $\lambda^{\ast} = \Omega_{2}$ is unique, achieving the minimum value of the objective function in (48a). Otherwise, the water level continues to increase. When $\lambda \in (\Omega_{2}, \Omega_{3})$, $Y^{\ast}_{2} = 1$, and $Y^{\ast}_{1} = Y^{\ast}_{3} = Y^{\ast}_{4} = 0$. If these optimal values satisfy the constraint in (49b), the water filling process halts. For this case, the optimal $\lambda$ is not unique, but all yield the same optimal objective function in (48a). If these optimal values do not satisfy the constraint in (49b), the water-filling process continues until the optimal value for the objective function in (48a) is obtained.

\section{Numerical Results}
\label{Section_simulation}
In this section, we conduct numerical studies on the approaches put forth in the preceding section. To be specific, we numerically exam the optimal solutions to the problem described in equations (48) and (30), respectively. The data model considered in the simulations is
\begin{align}
    {x}_{j,l} = \theta + {n}_{j,l},
    \tag{56}
\end{align}
where $\theta$ is a deterministic unknown parameter, and ${n}_{j,l}$ is Gaussian noise with standard normal distribution.
We assume that if ${x}_{j,l}$ is falsified by an adversary, the adversary employs a data injection attack to falsify ${x}_{j,l}$ to
\begin{align}
    {\tilde{x}}_{j,l} = \theta + \xi_{j,l} + {n}_{j,l},
    \tag{57}
\end{align}
where $\xi_{j,l}$ is an unknown attack parameter. Also, each IoT device employs an one-bit quantizer to convert ${x}_{j,l}$ to a quantized data ${u}_{j,l}$ before transmitting them to miners.

In Fig. 2, we consider the scenario where $\theta$ = 2, $\xi_{j,l}$ = 2.5 for all $j$ and $l$, and $L_0$ = 4. Thus, the choices of $L_a$ that the adversary can select are {1, 2, 3, and 4}. In addition, we choose $N$ = 3, $\mid\mathcal C_{0}\mid$ = 2, $\mid\mathcal C_{a}\mid$ = 1, ${p}_{\theta}=[p_{\theta,0}, p_{\theta,1}]$ = [0.02871656, 0.97128344], and $\tilde{p}_{\boldsymbol\eta} = [\tilde{p}_{\boldsymbol{\eta},0}, \tilde{p}_{\boldsymbol{\eta},1}]$ = [0.06680720
, 0.93319280]. In the simulation depicted in Fig. 2, we consider $L$ = 5, 6, 7, 8, 9, and 10. The corresponding $P{(L_a)}$ = [0.00243, 0.0081, 0.027, 0.09], [0.00499, 0.0149, 0.044, 0.1278], [0.0071, 0.0199, 0.0548, 0.148], [0.0088, 0.0235, 0.0617, 0.156], [0.010, 0.0261, 0.06645, 0.166], and [0.0111, 0.027997, 0.06972, 0.1715] respectively. Moreover, the optimal of (30) is obtained by using the gradient descent method with multiple starting points\cite{44}.
\begin{figure}[!h]
  \centering
    \includegraphics[width=8cm]{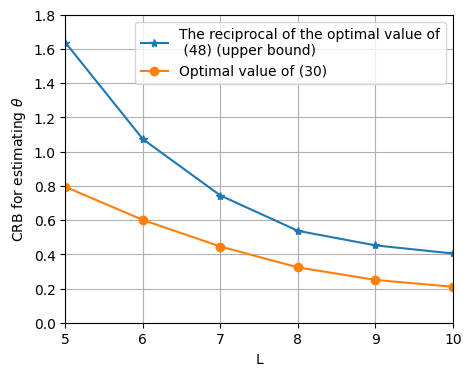}
  \caption{Comparison between the optimal value of (30) and the reciprocal of the optimal value of (48) for each $L$.}
  \centering
    \includegraphics[width=8cm]{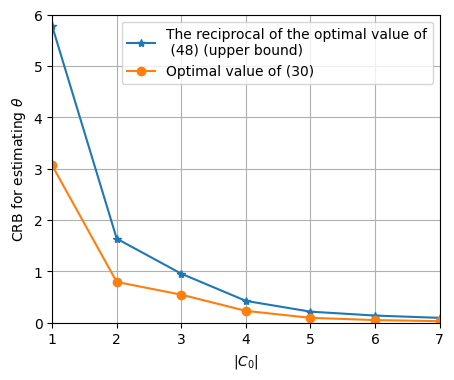}
  \caption{Comparison between the optimal value of (30) and the reciprocal of the optimal value of (48) for each $|\mathcal{C}_0|$.}
  \centering
    \includegraphics[width=8cm]{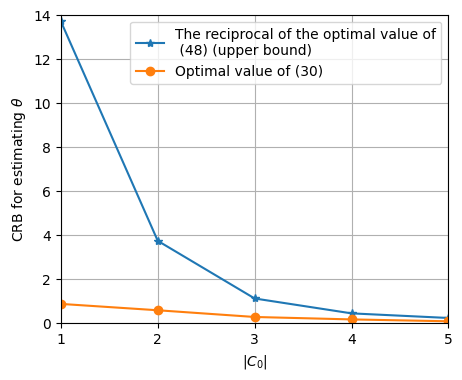}
  \caption{Comparison between the optimal value of (30) and the reciprocal of the optimal value of (48) for each $|\mathcal{C}_0|$.}
\end{figure}
It is seen from Fig. 2 that as $L$ increases from 5 to 10, both the optimal value of (30) and the reciprocal of the optimal value of (48) decrease. This trend arises because as $L$ increases, the number of data stored in the blockchain increases, leading to a better estimation performance. Furthermore, the numerical results in Fig. 2 corroborate our theoretical results that the reciprocal of the optimal value of (48) serves as a valid upper bound on the optimal value of (30).

In Fig. 3, we consider the same parameters as those considered in Fig. 2, and choose $L$ = 5 and $\mid\mathcal{C}_{a}\mid$ = 1, while increasing the number of honest IoT devices, $\mid\mathcal{C}_{0}\mid$. As $\mid\mathcal{C}_{0}\mid$ increases, the number of IoT devices, $N$, increases. It is seen from Fig. 3 that the optimal value of (30) always remains less than the reciprocal of the optimal value of (48) which agrees with our theoretical results. Furthermore, with increasing $\mid\mathcal{C}_{0}\mid$, the estimation performance tends to improve. This is due to the fact that as the number of honest IoT devices increases, the number of unattacked data increases, and hence the estimation performance improves.

In Fig. 4, we investigate the case where $N = 6$ and $L = 5$, while we vary the sizes of $\mathcal{C}_{0}$ and $\mathcal{C}_{a}$. Similar to Fig. 2 and Fig. 3, Fig. 4 demonstrates that the reciprocal of the optimal value of the relaxed optimization problem in (48) serves as an upper bound on that of (30). Additionally, as illustrated in Fig. 4, the larger the $\mid\mathcal{C}_{0}\mid$, the better the estimation performance. This is due to the fact that as $\mid\mathcal{C}_{0}\mid$ increases, the percentage of malicious IoT devices decreases, and therefore, the impact of malicious data on the estimation performance is diminished.

\section{Conclusions}\label{Section_conclusion}
In this paper, we consider distributed estimation in a BIoT under attacks that jointly exploit the vulnerabilities of both IoT devices and the blockchain employed in the BIoT. We adopt the CRB as the performance metric, and we aim to derive the estimation performance guarantee for the BIoT under attacks by maximizing the CRB over the attack domain. To achieve this, we develop a relaxation method to convert the original non-convex optimization problem into a convex problem. We also derive the analytic expression for the optimal solution to the relaxed convex problem. Our theoretical and numerical results demonstrate that the optimal value of the relaxed optimization problem can be used to serve as a valid estimation performance guarantee for the BIoT in the presence of attacks.


\appendices



%
%

\bibliography{ref}
\bibliographystyle{IEEEtran}
\end{document}